\documentclass[aps,preprint]{revtex4-1} 
\usepackage{graphicx}
\usepackage{graphics}
\expandafter\let\csname equation*\endcsname\relax 
\expandafter\let\csname endequation*\endcsname\relax 

\usepackage{amsmath}

\newcommand{\half}{\frac{1}{2}}

\newcommand{\eps}{\epsilon}

\newcommand\refeq[1]{(\ref{#1})}
\newcommand\reffig[1]{Figure~\ref{#1}}
\newcommand\of[1]{\left( #1 \right)}
\newcommand\sqof[1]{\left[ #1 \right]}

\begin{document}

\title{Relativistic Linear Restoring Force}
\author{D. Clark}
\author{J. Franklin}
\author{N. Mann}
\email{nmann@reed.edu}
\affiliation{Department of Physics, Reed College, Portland, Oregon 97202,
USA}

\begin{abstract}
We consider two different forms for a relativistic version of a linear restoring force.  The pair comes from taking Hooke's law to be the force appearing on the right of the relativistic expressions: $\frac{d p}{dt}$ or $\frac{dp}{d\tau}$ . Either formulation recovers Hooke's law in the non-relativistic limit.  In addition to these two forces, we introduce a form of retardation appropriate for the description of a linear (in displacement) force arising from the interaction of a pair of particles with a relativistic field.  The procedure is akin to replacing Coulomb's law in E\&M with a retarded form (the first correction in the full relativistic case).  This 
retardation leads to the expected oscillation, but with amplitude growth in both its relativistic and non-relativistic incarnations.
\end{abstract}

\maketitle

\section{Introduction}

By the end of a student's undergraduate career, the classic harmonic oscillator has appeared in almost every imaginable context.  We see it in introductory physics, as our first example of a spatially varying force, again in a sophomore-level physics class, as a vehicle for studying the Fourier Transform, and maybe again in a computational physics course, as a model for bonds connecting atoms.  Aside from its direct utility, the equation of motion associated with the harmonic oscillator occurs in a variety of settings, and most instructors refer us back to our earlier experience with masses connected by springs as a way to visualize widely disparate physical systems that have, at their base, the same fundamental equation and solution.   One reason for the central role of the harmonic force is its near-universal applicability -- for almost any potential that has a minimum, motion in the vicinity of that minimum is governed by a linear restoring force with a ``spring constant" related to the curvature of the potential at the minimum.

When we move to a relativistic description of simple harmonic motion, that universality is lost.  
There is an {\it a priori} ambiguity in the interpretation of a classical force $F$ in relativistic dynamics.  It could either appear as a component of a three-force, so that $\frac{dp}{dt} = F$, or it could be a component of a four-force, in which case $\frac{dp}{d\tau} = F$.  In E\&M, for example, the Lorentz force is an ``ordinary" three-force (see p. 519~\cite{Griffiths}).  For our non-relativistic $F = -k\, x$, we must first determine what type of formulation (three or four-force) to use.  
However, both of these formulations share the same behaviors in
the non-relativistic and ultra-relativistic limits.  Indeed, we could
argue that correct behavior in these limits is the minimal requirement
for a system to be considered a ``relativistic spring."  In fact, we
can define an infinite family of equations of motion with the right limiting behaviors to be associated with a relativistic form of Hooke's law.  Following~\cite{Dylan},  we will focus on the three/four-force pair associated with $F= -k \, x$.  We'll start by considering the relativistic dynamics of each of these.

If we want either form of the force to correspond, as in E\&M, to an interaction of particles with fields, we must introduce a retarded time evaluation.  We proceed to do this for a pair of particles, assuming that the governing field moves at speed $c$ in vacuum.  We assume the force depends linearly on the distance between the particles, taking into account the time of flight for the field linking the particles.  This process represents a minimal substitution, in which we effectively replace the instantaneous evaluation of displacement with a retarded-time evaluation.  For this modified form we find a physically implausible growth in oscillatory amplitude.  The pair of particles by themselves no longer constitute an isolated system, and unless we include contributions from the field, we appear to have lost conservation of energy.

This paper is meant to demonstrate the logic and physical checks that are common in high-energy and gravitational model building as applied to a toy system that is familiar from the undergraduate curriculum.  In the spirit of demonstration, we take a pair of simple choices (letting $-k \, x$ play the role of a component of a $3$ or $4$-force) and explore the physical implications of these two systems.

\section{Non-Relativistic Springs}\label{NRS}

Let's start by reviewing the standard problem and its solution.  For a mass $m$ attached to a spring with spring constant $k$ and equilibrium at zero, Hooke's law combined with Newton's second law give the equation of motion:
\begin{equation}\label{FNEOM}
\frac{d}{dt} \, \sqof{ m \, \dot x} = -k \, x.
\end{equation}
If we start with $x(0) = a$ and $\dot x(0) = 0$, then the solution is
\begin{equation}
x(t) = a \, \cos\of{\sqrt{\frac{k}{m}}\, t}.
\end{equation}
One immediate (special relativistic) problem with this solution is that the speed of the mass,
\begin{equation}
\dot x(t) = a \, \sqrt{\frac{k}{m}} \, \sin\of{\sqrt{\frac{k}{m}} \, t},
\end{equation}
has a maximum value of $v_{\hbox{\tiny max}} = a \, \sqrt{\frac{k}{m}}$.  This value can be made greater than $c$ for some choice of $a$.

\section{Relativistic Linear Displacement Force}

Suppose we start by considering the simplest possible equation of motion, given by $\frac{dp}{dt} = -kx$.  We'll take $p$ to be the relativistic momentum, giving 
\begin{equation}\label{N2SR}
\frac{d}{dt} \, \sqof{\frac{m\, \dot x}{\sqrt{1 - \frac{\dot x^2}{c^2}}}} = -k\, x.
\end{equation}
From this form we can immediately make a qualitative observation about the extreme relativistic oscillatory behavior.  Assume the same initial conditions as above.  For $a$ large, the force is large, and the mass undergoes rapid acceleration from rest.  But the maximum achievable speed is $c$ (this is enforced by the form of the relativistic momentum appearing in~\refeq{N2SR}), so the acceleration must die off as the mass approaches speed $c$ -- the larger the initial extension, the faster the value of $c$ is attained.  Once the mass has travelled to $-a$, it is at rest, and the process begins again in the opposite direction.  Taken to the extreme, the resulting motion looks like a sawtooth, with the mass moving at a constant speed $c$ from $a$ to $-a$, instantaneously reversing direction, and again moving at speed $c$ from $-a$ to $a$.  A physical manifestation of this motion would be light bouncing back and forth between two mirrors.

In fact, any sensible ``relativistic version'' of a spring must have this behavior in the extreme relativistic limit, in addition to behaving like a traditional non-relativistic spring in the opposite limit.  Aside from details, we have a clear picture of what must happen: a smooth transition from the non-relativistic sinusoidal motion to the extreme sawtooth motion.  This is already an interesting prediction, since we know that in the non-relativistic limit the period of the motion is given by $T = 2 \, \pi \, \sqrt{\frac{m}{k}}$, a quantity that is independent of the mass's initial conditions.  In the extreme relativistic limit (where $a$ is large), the period of the motion is $T = 4\, a/c$, and so is governed {\it entirely} by the initial extension, with no spring information involved.  One interesting consideration is the manner (as a function of $a$) in which the system ``loses" its knowledge of the spring.  We will find that the two forms for our equation of motion, $\frac{dp}{dt} = -k\, x$ and $\frac{dp}{d\tau} = -k\, x$, give different results.

\subsection{Hooke's Law Forces}

Consider a diametric hole drilled through a uniformly charged sphere.  A charged particle moving in the hole experiences a linear restoring force.  The equation of motion is:
\begin{equation}\label{Mforce}
\frac{d p}{dt} = \frac{d}{dt} \, \sqof{ \frac{m \, \dot x}{\sqrt{1 - \frac{\dot x^2}{c^2}}}} = -k \, x,
\end{equation}
where $k$ is fixed by E\&M constants.  We can isolate acceleration on the left, giving us an equation that can be compared with its Newtonian counterpart:
\begin{equation}\label{EOMrev1}
m \, \ddot x = -k \, x \, \of{1 - \frac{\dot x^2}{c^2}}^{3/2}.
\end{equation}
Starting from rest with extension $a$ it is possible to solve for $x(t)$ in terms of incomplete elliptic functions~\cite{MacColl}.

In~\refeq{Mforce}, we took $-k \,x$ to be a component of the three-force, so that $\frac{dp}{dt} = -k \, x$.  This formulation makes sense for a particle moving in an external harmonic potential, one in which the constant $k$ is fixed in the ``lab" frame.  But, we could also let $k$ remain fixed in the rest frame of the moving particle -- in this case, $\frac{dp}{d\tau} = -k\, x$ is the relevant equation of motion, with $-k \, x$ appearing as a component of the four-force.  We could engineer this case by changing the density $\rho(t)$ in our uniformly charged sphere so as to achieve constant $k$ in the instantaneous rest frame of the particle \footnote{This type of linear restoring force, with constant $k$ defined in the particle rest frame, can also show up in geodesic motion for background metrics that are not flat~\cite{Franklin}.}.  Physical configuration aside, if we move the non-relativistic potential $\half \, k \, x^2$ to the Lorentz scalar potential $\half \, k \, x^\mu \, x_\mu$ (for $x_\mu \dot = (c \, t \, \,  \, , \, \, x)^T$), then the (spatial) equation of motion would read: 

\begin{equation}
\frac{dp}{d\tau} = \frac{1}{\sqrt{1 - \frac{\dot x^2}{c^2}}}\, \frac{dp}{dt}  = \frac{1}{\sqrt{1 - \frac{\dot x^2}{c^2}}}\, \frac{d}{dt}\, \sqof{\frac{m \, \dot x}{\sqrt{1 - \frac{\dot x^2}{c^2}}}}    = -k \, x.
\end{equation}
The form equivalent to~\refeq{EOMrev1} is
\begin{equation}\label{FIIEOM}
m \, \ddot x = - k \, x \, \of{1 - \frac{\dot x^2}{c^2}}^2.
\end{equation}

To express both of these options in coordinate time, where $\frac{d p}{d t}$ is the relevant quantity, we will write these two equations as $\frac{d p}{dt} = F_t$ and  $\frac{d p}{dt} = F_\tau$, with 
$F_t \equiv -k \, x$ the three-force formulation and  $F_\tau \equiv -k \, x \, \sqrt{1 - \frac{\dot x^2}{c^2}}$ the four-force formulation.

\section{Quantitative Comparison}\label{Quant}

We now want to explore the details of our two dynamical systems.  This is made clearer by rendering all of our equations dimensionless.  To do so we use the length scale $a$ set by initial conditions and the natural time scale $\sqrt{m/k}$ (from Section \ref{NRS}).  Let $x = a \, q$ and $t = \sqrt{m/k} \, s$ so that $x(0) = a$ becomes $q(s=0) = 1$.   We also define the ratio of initial potential energy to rest energy, $\sigma \equiv \frac{\half \, k \, a^2}{m \, c^2}$, a parameter that governs the size of the relativistic correction.  Then our equations of motion become, in order (i.e.~\refeq{FNEOM}, ~\refeq{EOMrev1}, ~\refeq{FIIEOM}):
\begin{equation}\label{eomfinal}
\begin{aligned}
q_N''(s) &= - q_N(s) \\
q_t''(s) &= -q_t(s) \, \of{1 -2 \, \sigma\,  q'_t(s)^2}^{3/2} \\
q_{\tau}''(s) &= -q_\tau(s) \,\of{1 - 2 \, \sigma \, q'_\tau(s)^2}^2.
\end{aligned}
\end{equation}

\begin{figure}[htbp] 
   \centering
   \includegraphics[width=4in]{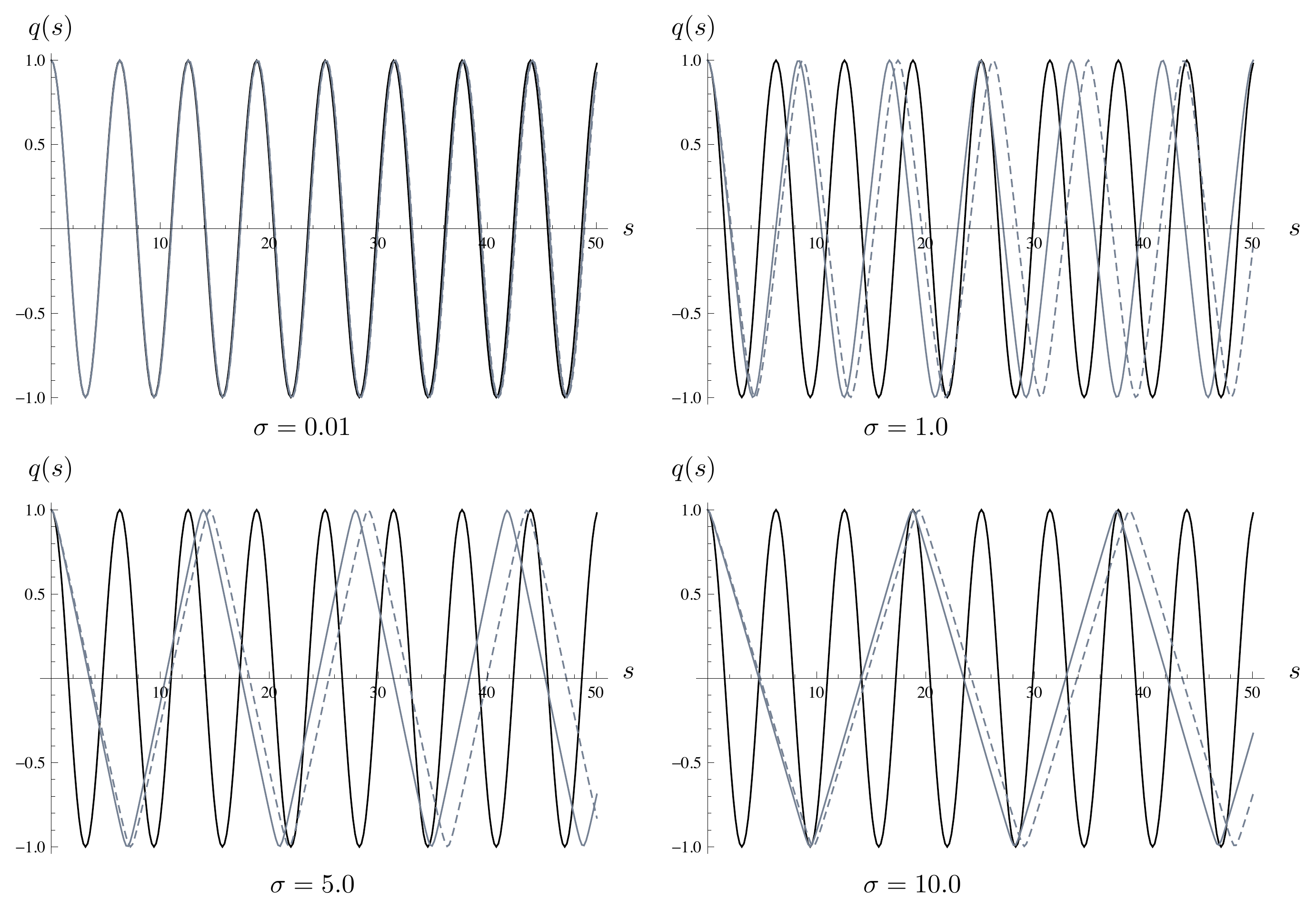} 
   \caption{The non-relativistic motion is shown in black ($q_N(s)$), and relativistic motion under the influence of $F_t$ and $F_\tau$ are shown in gray ($q_t(s)$) and dashed gray ($q_{\tau}(s)$) respectively.  The motion is displayed for increasing energy (ratio) $\sigma$.}
   \label{fig:motion}
\end{figure}

We can solve all three equations in~\refeq{eomfinal} for various $\sigma$ -- the result of that numerical calculation~\footnote{We used a fixed-step Runge Kutta method.  The step-size was determined by requiring that the relativistic energy, in the case of $F_t$, be conserved at one part in $10^{10}$.} is shown in~\reffig{fig:motion}.
Referring to the figure we see that for small $\sigma$, the three separate equations of motion have very similar solutions.  This confirms that both relativistic forms revert to the non-relativistic solution for small $\sigma$, which is also clear from the expressions in~\refeq{eomfinal}.  As $\sigma$ becomes large, both relativistic forms move towards a sawtooth, as they must.

As expected, the low energy limits of the two trajectories coincide, and the high energy limits, even at $\sigma = 10$, are beginning to converge.  However,~\reffig{fig:motion} does not present a good way to explore the intermediate regime.   The solutions to~\refeq{eomfinal} are periodic for all $\sigma$, and we know the period of motion in both limits:  the Newtonian limit  $2 \, \pi$ is immediately obvious, and the sawtooth period is $4 \, \sqrt{2 \, \sigma}$.   We can then compare $T_t(\sigma)$ and $T_\tau(\sigma)$ (the periods of the motion under the influence of $F_t$ and $F_\tau$ respectively) by direct integration.

The equations in~\refeq{eomfinal} are all of the form
\begin{equation}\label{gencase}
q''(s) = -q \, \of{1 - 2 \, \sigma \, q'^2}^p
\end{equation}
for $p = \{0, 3/2, 2\}$.  This general form can be integrated once to give 
\begin{equation}\label{Aconst}
A = \frac{\of{1 - 2 \, \sigma \, q'^2}^{1-p}}{4 \, \sigma\, \of{p-1}} + \half \, q^2 
\end{equation}
for constant $A$.  In the case of $p = 0$, the constant $A$ is the non-relativistic energy (with a constant offset), and for $p=3/2$, it is the relativistic energy.   For $p=2$,~\refeq{Aconst} implies a conserved Hamiltonian, however $A$ cannot be interpreted as the energy 
\footnote{That Hamiltonian has equations of motion that correctly reproduce~\refeq{gencase}, and so it is the generator of time-translation, and a constant.  This provides a formal definition of energy for all values of $p$, even if it does not correspond to one of the familiar cases.}.  If we use the initial conditions ($q(0) = 1$, $q'(0) = 0$, appropriate for a particle starting from rest) to fix the constant, then we can solve for $q'(s)$ in terms of $q(s)$, yielding
\begin{equation}\label{qpofp}
q'(s) = \pm \frac{1}{\sqrt{2 \, \sigma}} \, \of{1 - \sqof{1 + 2 \, \sigma \, (p-1) \, \of{1 - q^2}}^{\frac{1}{1-p}}}^{1/2}.
\end{equation}

Note that one can also consider~\refeq{gencase}-\refeq{qpofp} for other values of $p$ as they define an infinite family of relevant effective forces.  Any suitable relativistic model must have the same non-relativistic and ultra-relativistic limits (as these do), but beyond that, there is  freedom.  In addition to the two specific cases of interest to us, a model resulting in $p=1$ was developed from material considerations in \cite{Gron}.

\begin{figure}[htbp]
   \centering
   \includegraphics[width=4in]{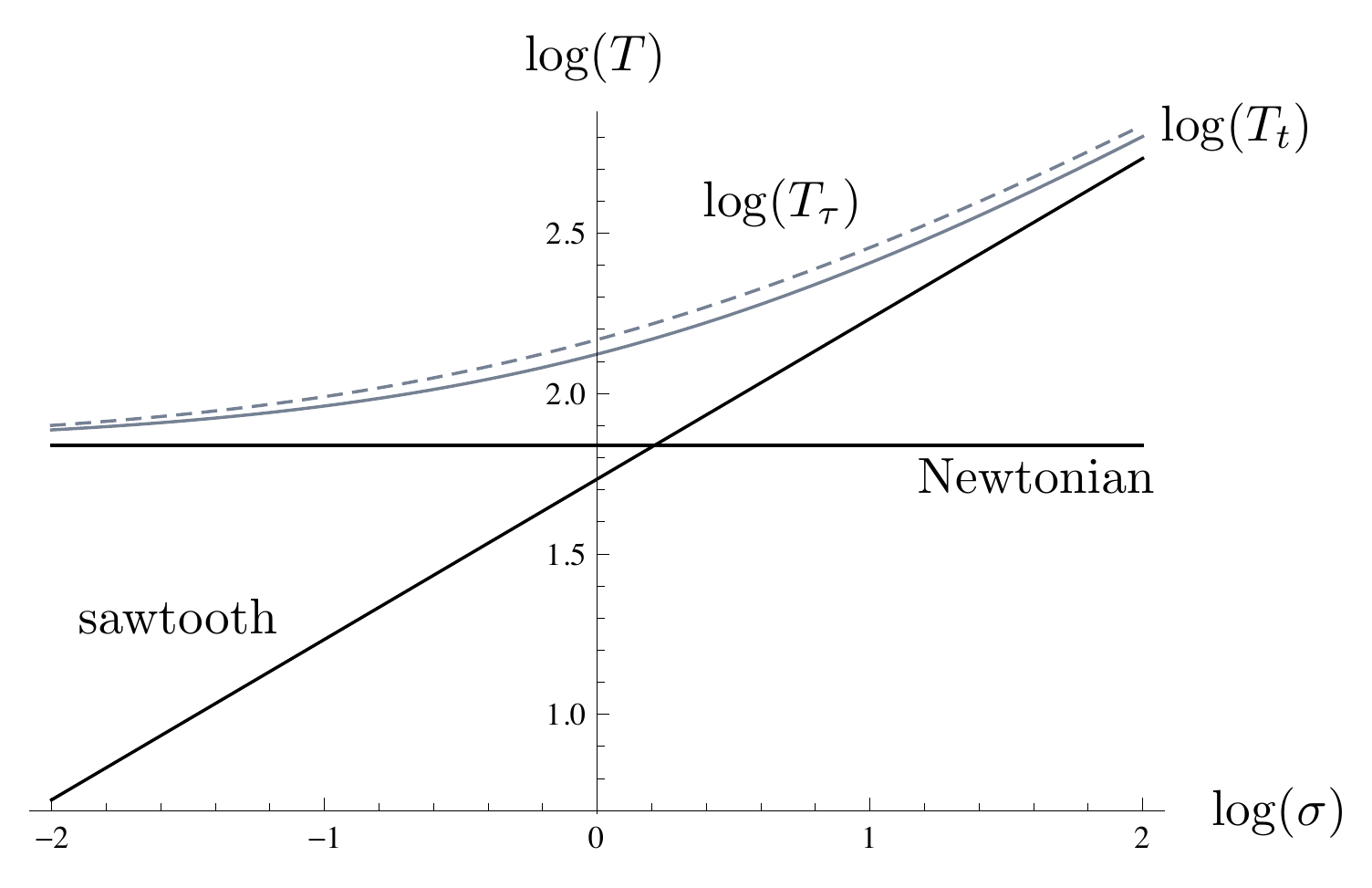} 
   \caption{The period of motion as a function of $\sigma$ for the solutions to~\refeq{eomfinal} -- here, we use $\log(\sigma)$ to capture a wide range of behavior.  The non-relativistic and relativistic limits are shown in black, the periods under the influence of forces $F_t$ and $F_\tau$ are shown in gray and dashed gray.}
   \label{fig:Periods}
\end{figure}

Using the general expression~\refeq{qpofp}, we can integrate over a quarter of a period with definite sign.   For $q_t$ ($p = 3/2$), this gives:
\begin{equation}
T_t(\sigma) =4\, \sqrt{\frac{2}{2 + \sigma}} \, \sqof{(2 + \sigma)\,  E(\sigma/(2 + \sigma))- K(\sigma/(2 + \sigma))}
\end{equation}
and for $q_{\tau}$ (with $p =2$) we have:
\begin{equation}
T_{\tau}(\sigma) = 4 \, \sqrt{1 + 2 \, \sigma} \, E(2 \, \sigma/(1+ 2 \, \sigma)).
\end{equation}
In both expressions, $K$ and $E$ are the complete elliptic integrals of the first and second kinds~\footnote{These are defined to be
\begin{equation*}
\begin{aligned}
K(\alpha) &\equiv  \int_0^{\frac{\pi}{2}} \frac{1}{\sqrt{1 - \alpha \, \sin^2\phi}} \, d\phi  \\
E(\alpha) &\equiv  \int_0^{\frac{\pi}{2}}  \sqrt{1 - \alpha \,  \sin^2\phi} \, d\phi
\end{aligned} 
\end{equation*}}.  These two periods, together with the limiting cases (the sinusoidal Newtonian one, and relativistic sawtooth), are plotted in~\reffig{fig:Periods}.  We see that the period $T_\tau(\sigma)$ diverges from the non-relativistic period more quickly, and converges to the relativistic limit more slowly, than $T_t(\sigma$).  Hence the transition between the two limits is more gradual for $T_\tau(\sigma)$ than for $T_t(\sigma)$.  The period for any value of $p$ should look similar to these, with a different transition from one limit to the other.

\section{Information Speed}

\begin{figure}[htbp] 
   \centering
   \includegraphics[width=2in]{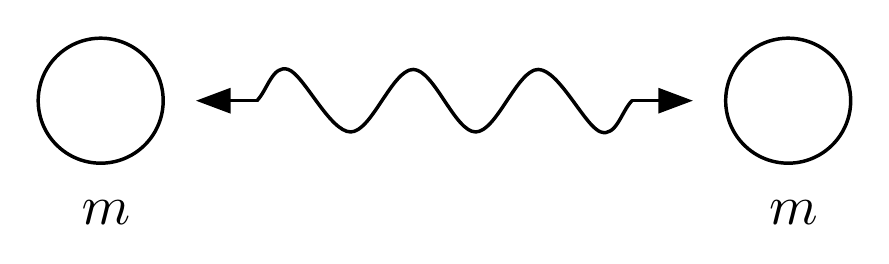} 
   \caption{A field mediates the interaction between two particles of equal mass, producing a force $F_\tau$.}
   \label{fig:twoparticles}
\end{figure}

There is another element to relativistic systems:  the question of causality.   Our model for exploring this will be a linear (in displacement) restoring force with a finite speed of propagation for the interaction.  For concreteness, we will take the form $F_\tau$, although  the qualitative features of the dynamics produced by $F_t$ and $F_\tau$ are similar~\cite{Dylan} \footnote{In the specific physical model motivating  $F_t$, as presented here, the force is determined by a static external electromagnetic configuration and therefore there is no issue with the speed of information.}.  We will suppose that underlying the production of $F_\tau$ is a field that mediates the interaction between two particles as in~\reffig{fig:twoparticles}.  Given material considerations, it would be natural to imagine that field moving between the particles with some fundamental speed $v$, determined by the properties of the supporting medium.  

If we have two equal masses that interact via $F_\tau$, and we do {\it not} consider a time-of-flight correction, then the equations for the left and right masses should be:
\begin{equation}
\begin{aligned}
F_\ell(t) &= k \, \sqof{ x_r(t) - x_\ell(t)} \, \sqrt{1 - \frac{\dot x_\ell(t)^2}{c^2}} \\
F_r(t) &= -k \, \sqof{ x_r(t) - x_\ell(t)} \, \sqrt{1 - \frac{\dot x_r(t)^2}{c^2}}
\end{aligned}
\end{equation}
where each mass experiences a force that depends on the instantaneous location of the other.

In order to introduce dependence on an information propagation speed $v$, we will evaluate the force on the right hand particle using its location at time $t$, but use an earlier position for the left-hand mass: $x_\ell(t_r)$, where $t_r$ is the retarded time defined by the condition:
\begin{equation}\label{vttr}
v \, (t - t_r) = | x_r(t) - x_\ell(t_r)|.
\end{equation}
Assuming the masses are given symmetric initial conditions (so that the left mass starts at $x_\ell(0) = -a$ with $\dot x_\ell(0) = 0$, and the right mass has $x_r(0) = a$ with $\dot x_r(0) = 0$), we can write
\begin{equation}\label{Fret}
\begin{aligned}
F_\ell(t) &= k \, \sqof{ x_r(t_r) - x_\ell(t)}\, \sqrt{1 - \frac{\dot x_\ell(t)^2}{c^2}} \\
F_r(t) &= -k \, \sqof{ x_r(t) - x_\ell(t_r)}\, \sqrt{1 - \frac{\dot x_r(t)^2}{c^2}}.
\end{aligned}
\end{equation}
This move is analogous to taking the Coulomb force between two particles in one dimension (labelled left and right as above):
\begin{equation}
F_r = \frac{q_1 \, q_2}{4 \, \pi \, \eps_0 \, |x_r(t) - x_{\ell}(t)|^2},
\end{equation}
and replacing the difference $x_r(t) - x_\ell(t)$ with $x_r(t) - x_{\ell}(t_r)$ (with $v = c$ in~\refeq{vttr}).  That replacement would yield the correct force up to corrections of order $\dot x_r/c$ that come from the actual field of a moving point particle.  Motivated by simplicity, we will start
with~\refeq{Fret} and see what solutions emerge.

If we begin with the stated (symmetric) initial conditions and define $x(t) \equiv  x_r(t) =-x_\ell(t)$, then the single equation of motion together with the retarded time definition is
\begin{equation}\label{relmomeom}
\begin{aligned}
p(t) &= \frac{m \, \dot x(t)}{\sqrt{1 - \frac{\dot x(t)^2}{c^2}}} \\
\frac{d p(t)}{d t} &= -k \, \sqof{x(t) + x(t_r)}\, \sqrt{1 - \frac{\dot x(t)^2}{c^2}} \\
v \, \of{t - t_r} &= |x(t) + x(t_r)|.
\end{aligned}
\end{equation}
In order to get the right-hand side of the equation for $\frac{dp}{dt}$ in terms of $p$, we must invert the first equation in~\refeq{relmomeom} to find $\dot x$ as a function of $p$.  When we do this, and
use the same dimensionless factors as above, with the introduction of $p =(a \sqrt{m \, k}) \, w$ for dimensionless momentum $w$, we have:
\begin{equation}\label{relspring}
\begin{aligned}
\frac{dq(s)}{ds} &= \frac{w(s)}{\sqrt{1 + 2 \, \sigma \, w(s)^2}} \\
\frac{d w(s)}{d s} &= - \frac{q(s) + q(s_r)}{\sqrt{1 + 2 \, \sigma \, w(s)^2}} \\
\frac{v}{c} \, (s - s_r) &= \sqrt{2 \, \sigma} \, |q(s) + q(s_r)|.
\end{aligned}
\end{equation}
The initial conditions are $w(0) = 0$, $q(0) =1$, and we assume (for the purposes of computing $s_r$) that these values hold for all $s \le 0$.  It is clear that in the limit $\sigma \longrightarrow 0$, we recover the familiar instantaneous displacement, $s = s_r$, and the correct non-relativistic equations of motion.

The dimensionless factor $v/c$ gives us an additional knob -- we can elect to set $v \ll c$ while maintaining finite relativistic energies (the $v/c \longrightarrow 0$ limit functions, in the retardation condition, much like the $\sigma \longrightarrow \infty$ limit).  The other limit, $v = c$ is the one on which we will focus on the grounds that:  1.\ we have no physical model with which to fix $v$, and 2.\ using $v<c$ allows multiple solutions to~\refeq{vttr} (since the particles themselves can travel faster than $v$, allowing multiple points of causal contact).  This could lead to interesting dynamics, however it is not clear how to weight the effects of the multiple solutions.  Inspired again by electrodynamics, we will make the simplifying assumption that our field travels at the speed of light.

We are ready to numerically solve the set~\refeq{relspring} for arbitrary $\sigma$.  
The method we use to solve the equations of motion in~\refeq{relspring} is basically velocity Verlet~\cite{MMOL}, a finite difference approach where we use a fixed grid $s_j = j \, \Delta s$, and approximate the ODEs in~\refeq{relspring} at these grid points.   For the retarded time condition in~\refeq{relspring}, we use a root-finding routine (bisection) adapted to the grid that will allow us to solve for $s_r$ at $s_j$ up to errors of size $\Delta s/2$.  The scheme is easy to develop and we present it briefly in the Appendix, so that the interested student could implement it.

\begin{figure}[htbp] 
   \centering
   \includegraphics[width=3in]{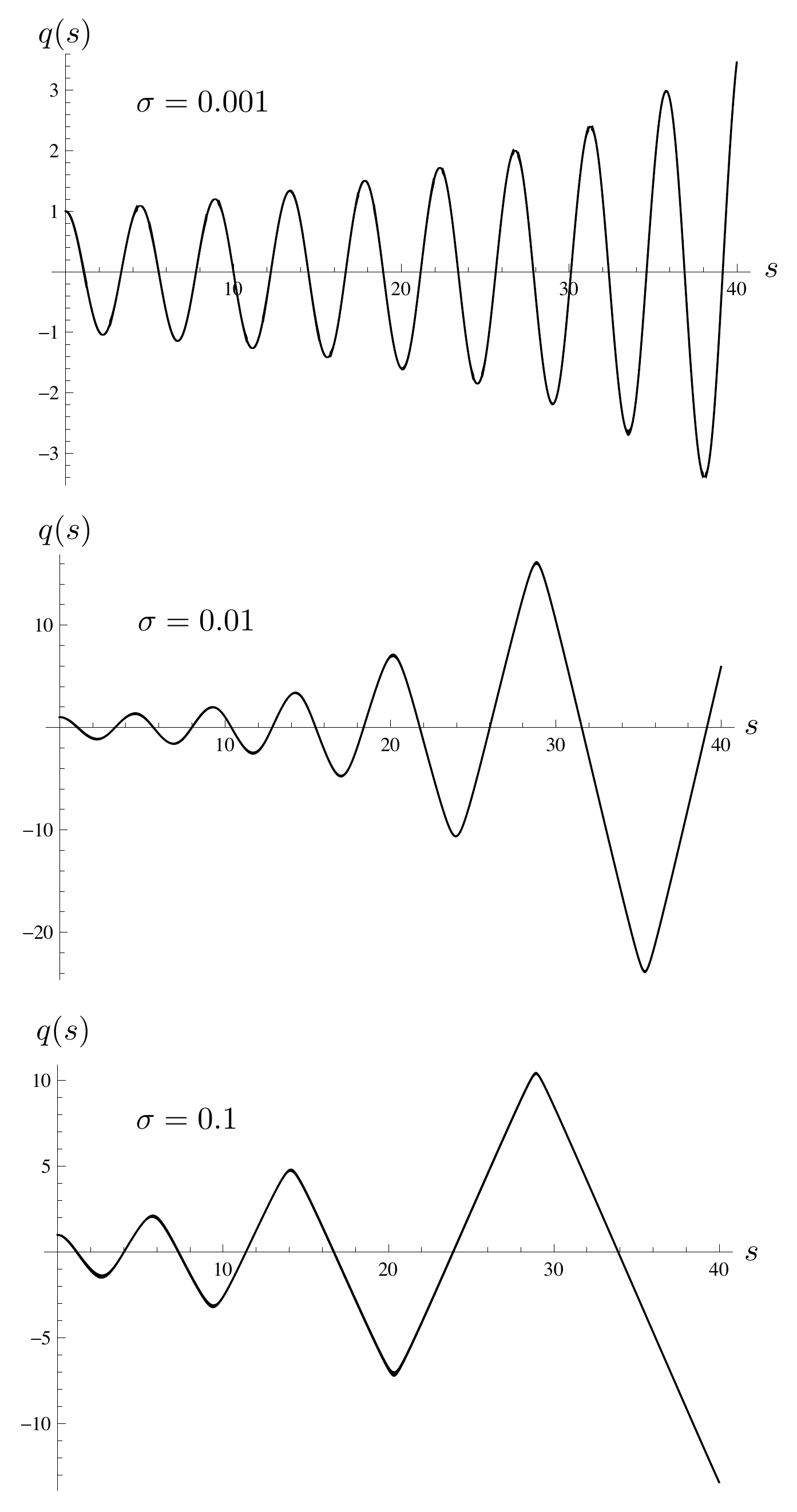} 
   \caption{The numerical trajectories approximating solutions to~\refeq{relspring}, shown for increasing value of $\sigma$.  In all cases, we use a step size, in~\refeq{update} of $\Delta s = 0.01$.}
   \label{fig:sigmatraj}
\end{figure}

When the method~\refeq{update} is iterated, we generate approximations to the delay differential equation defined by~\refeq{relspring}, and the trajectories for a few values of $\sigma$ are shown in~\reffig{fig:sigmatraj}.
What we see there is amplitude growth: as time goes on, it looks as if energy is being added to the system.  Since we are using the relativistic second law, that additional energy makes the trajectory look more and more like a sawtooth.   The result is quite striking, and given that it is a numerical solution, we would like to carry out some analytical tests to confirm the basic behavior.  We will consider two checks, both appropriate for $\sigma$ small, where exact results are obtainable.  Both of these apply to the relativistic case as well.

The growth behavior of $q(s)$ is evident even in the non-relativistic limit with $\sigma $ small, where its source is relatively easy to trace.  The equation of motion becomes
\begin{equation}
\frac{d^2 q}{d s^2} = -\sqof{q(s) + q(s_r)}
\end{equation}
in our dimensionless variables -- if $\sigma$ is very small, then $s - s_r \sim \sqrt{\sigma}$ is close to zero.  We will approximate this small difference with a constant $0 < \eps \ll 1$ (a delay $>1$ would occur only if we allowed the masses to move with speeds greater than $c$).  We can then expand the equation of motion to get
\begin{equation}
\frac{d^2 q}{d s^2} = -\sqof{q(s) + q(s - \eps)} \approx -2 \, q(s)  + \eps \, q'(s),
\end{equation}
and see that it has an ``anti-damping" term $\eps \, q'(s)$.  The solution is
\begin{equation}
\begin{aligned}
q(s) &= \frac{e^{\half \, \eps \, s}}{\alpha} \, \sqof{\alpha \, \cos\of{\half \, \alpha \, s} - \eps \, \sin\of{\half \, \alpha \, s}},\\
 \alpha &\equiv \sqrt{8 - \eps^2}.
\end{aligned}
\end{equation}
In this form it is clear that for a constant small delay the solution grows with time (at least for small $s$).

\begin{figure}[htbp] 
   \centering
   \includegraphics[width=3.5in]{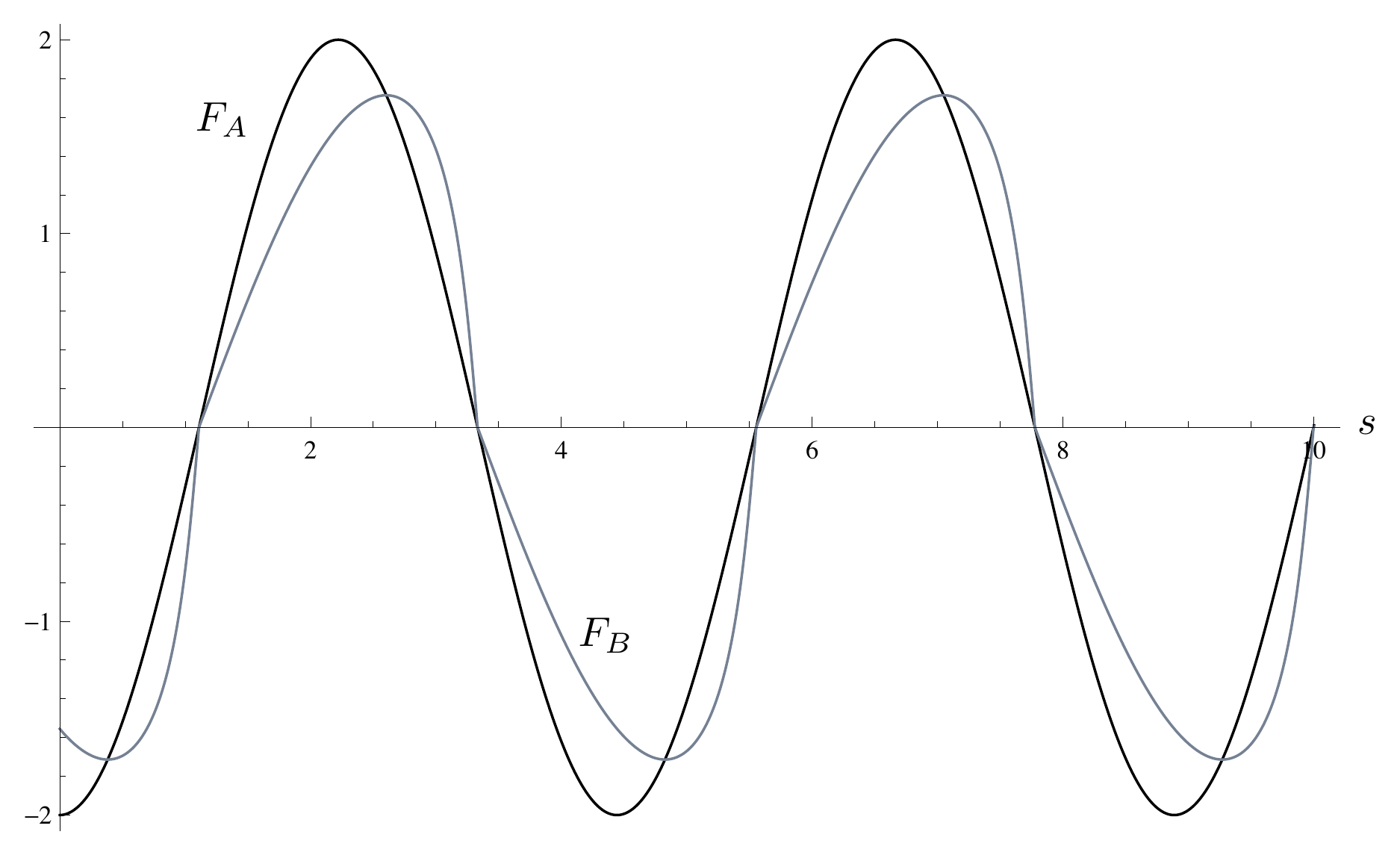} 
   \caption{The equal time spring force, $F_A$ (in black), and the retarded time spring force, $F_{B}$ (in gray) for a sinusoidal trajectory.}
   \label{fig:relForce}
\end{figure}

As a further check, we can take a continuous trajectory, $q(s) = \cos(\sqrt{2} \, s)$ (the solution to the non-relativistic problem with instantaneous forcing), and calculate both the equal-time spring force, $F_A \equiv -2 \, q(s)$, and the retarded time spring force, 
$F_{B} \equiv -\sqof{q(s) + q(s_r)}$, by solving the retardation condition for this analytic function.  The result is shown in~\reffig{fig:relForce} where we use $\sigma = 0.1$.  What we see is that the retarded force is large (relative to $F_A$) as the particle nears equilibrium, so that the mass gets a significant kick near $q = 0$.  Then the retarded force diminishes, and the particle continues with less deceleration than it would feel under the influence of $F_A$.  The process repeats; each time the particle nears equilibrium $F_{B}$ gets large compared to $F_A$ and then eases off.  The series of kicks, followed by decreased deceleration force, leads to an overall increase in amplitude.

Physically, the growth must be caused by whatever is mediating the interaction -- in E\&M we have fields to transmit the effect of one particle on another, and those fields carry energy.  
We have no model for the corresponding mediating ``field" in the retarded spring setting, so we have necessarily excluded its contribution to the total energy -- we then have an ``open" system, which doesn't conserve energy.

\section{Conclusion}
The description of a relativistic oscillatory system is less universal than its non-relativistic counterpart.  The immediate complication is the identification of $-k \, x$ as a component of the ordinary, three-force, or the Minkowski four-force that provides a starting point. 
 If a fixed electrostatic source is generating a harmonic potential, then the force we use in the relativistic Newton's second law must be
\begin{equation}
F_t = -k \,  x
\end{equation}
where $k$ is a constant (associated with the electrostatic sources) and $x$ is the position of a particle in the lab.  But we could also imagine $-k \, x$ as appearing in $\frac{d p}{d\tau} = -k \, x$, leading to 
\begin{equation}
F_\tau = -k \, x \, \sqrt{1 - \frac{\dot x^2}{c^2}}.
\end{equation}
We see that each is a reasonable candidate if we were to introduce a ``Hooke's law force" in the relativistic form of Newton's second law (and there are many others defined by~\refeq{gencase}).  Both forces lead to the same limiting behavior in the low and high energy limits, while differing in the intermediate region.  We can track these changes explicitly by examining the change in period as a function of $\sigma$, the ratio of initial potential energy to rest energy.

We introduced a time of flight delay in the separation $x$ appearing in $F_\tau$ in the simplest possible way.  The resulting retarded force causes amplitude growth which seems non-physical.  In a closed dynamical setting of particles generating and responding to fields, the force we have introduced would lead to energy loss by the field, but we have no way of performing the accounting that would balance the total energy of the system (lacking the field portion of the story).

Aside from developing some effective field with its own dynamics, it would be interesting to introduce damping in the retarded setting to try to tame the amplitude increase (that damping could come, for example, from radiation reaction if the particles are charged).  
\acknowledgements{The authors thank D. Griffiths for commentary on this article.}

\appendix
\section{The Numerical Method}

Our goal is to discretize the dimensionless equations of motion~\refeq{relspring}, and develop a recursion relation that allows us to move forward in time from the initial conditions.
Let $q(s_j) = q_j$ and $w(s_j) = w_j$, then from Taylor expansion we know that
\begin{equation}\label{dqdiscrete}
\frac{d q(s)}{d s} \biggr\vert_{s = s_j} \approx \frac{q_{j+1} - q_{j-1}}{2 \, \Delta s},
\end{equation}
and similarly for $\frac{d w}{ds}$.   We proceed to define an update by replacing derivatives in~\refeq{relspring} with finite approximations, obtaining
\begin{equation}\label{update}
\begin{aligned}
q_{j+1} &= q_{j-1} + \frac{2 \, \Delta s \, w_j }{ \sqrt{ 1 + 2 \, \sigma \, w_j^2}} \\
w_{j+1} &= w_{j-1} - \frac{2 \, \Delta s \, k \, \sqof{q_j + q(s_r)} }{ \sqrt{{1 + 2 \, \sigma \, w_j^2}}} \\
 \of{s_j - s_r} &= \sqrt{2 \, \sigma} \, |q_j + q(s_r)|.
\end{aligned}
\end{equation}
Given starting values $q_0$ and $w_0$, we can use~\refeq{update} to propagate the solution forward in time.  At each step, we must solve the retarded time condition by finding $k$ such that
\begin{equation}\label{discretesr}
 \of{s_j - k \, \Delta s} = \sqrt{2 \, \sigma} \, |q_j + q_k|
\end{equation}
using the set of positions $\{q_\ell\}_{\ell =0}^{j-1}$.  In practice, we cannot achieve equality in~\refeq{discretesr} using our fixed grid, so we take the $k$ that comes closest to equality.
That rounding introduces error $\Delta s$ in the determination of the retarded time, and hence $\Delta s$ in the evaluation of the force associated with the position at the retarded time.    
Then the update for $w_{j+1}$ is accurate up to errors that go like $\Delta s^2$ per step (since the retarded evaluation $q(s_r)$ appears with a factor of $\Delta s$ in front of it in~\refeq{update}).

\end{document}